\documentclass[twocolumn,apj]{emulateapj} 
\usepackage{apjfonts}
\newcommand{\rotate}{}
\usepackage{lscape}

\shorttitle{Radio Galaxy 3C\,79} 
\shortauthors{Fu and Stockton}
\journalinfo{accepted by the Astrophys. J.}
\submitted{Received 2007 October 10; accepted 2007 December 26}

\newcommand{\kms}{km s$^{-1}$}
\newcommand{\cc}{cm$^{-3}$}
\newcommand{\msun}{$M_{\odot}$}
\newcommand{\zsun}{$Z_{\odot}$}
\newcommand{\lothree}{$L_{\rm[O\,III]}$}
\newcommand{\othree}{[O\,{\sc iii}]}
\newcommand{\otwo}{[O\,{\sc ii}]}
\newcommand{\none}{[N\,{\sc i}]}
\newcommand{\ntwo}{[N\,{\sc ii}]}

\newcommand{\hetwo}{He\,{\sc ii}}
\newcommand{\stwo}{[S\,{\sc ii}]}
\newcommand{\eg}{e.g.,}
\newcommand{\ie}{i.e.,}
\newcommand{\hst}{{\it HST}}
\newcommand{\sersic}{S\'{e}rsic}
                                                                                
\begin{document}
\title{The Host Galaxy and The Extended Emission-Line Region of The Radio Galaxy 3C\,79\altaffilmark{1}}
\author{Hai Fu and Alan Stockton}
\affil{Institute for Astronomy, University of Hawaii, Honolulu, HI 96822; fu@ifa.hawaii.edu, stockton@ifa.hawaii.edu}

\altaffiltext{1}{
Based in part on observations obtained at the Gemini Observatory,
which is operated by the Association of Universities for Research in
Astronomy, Inc., under a cooperative agreement with the NSF on behalf of
the Gemini partnership: the National Science Foundation (United States),
the Particle Physics and Astronomy Research Council (United Kingdom),
the National Research Council (Canada), CONICYT (Chile), the Australian
Research Council (Australia), CNPq (Brazil) and CONICET (Argentina).
Gemini Program ID: GN-2006B-C-3.
Some of the data presented herein were obtained at the W.M. Keck Observatory, which is operated as a scientific partnership among the California Institute of Technology, the University of California and the National Aeronautics and Space Administration. The Observatory was made possible by the generous financial support of the W.M. Keck Foundation.
Based also in part on observations made with the NASA/ESA Hubble Space Telescope, obtained from the Data Archive at the Space Telescope Science Institute, which is operated by the Association of Universities for Research in Astronomy, Inc., under NASA contract NAS 5-26555. 
}

\begin{abstract}
We present extensive ground-based spectroscopy and \hst\ imaging of
3C\,79, an FR\,II radio galaxy associated with a luminous extended
emission-line region (EELR). Surface brightness modeling of an
emission-line-free \hst\ $R$-band image reveals that the host galaxy is
a massive elliptical with a compact companion 0\farcs8 away and 4
magnitudes fainter. The host galaxy spectrum is best described by an
intermediate-age (1.3 Gyr) stellar population (4\% by mass),
superimposed on a 10 Gyr old population and a power law
($\alpha_{\lambda} = -1.8$); the stellar populations are consistent with
super-solar metallicities, with the best fit given by the 2.5 \zsun\
models. We derive a dynamical mass of $4\times10^{11}$ \msun\ within the
effective radius from the velocity dispersion. The EELR spectra clearly
indicate that the EELR is photoionized by the hidden central engine.
Photoionization modeling shows evidence that the gas metallicity in both
the EELR and the nuclear narrow-line region is mildly sub-solar ($0.3 -
0.7$ \zsun) --- significantly lower than the super-solar metallicities
deduced from typical active galactic nuclei in the Sloan Digital Sky
Survey. The more luminous filaments in the EELR exhibit a velocity
field consistent with a common disk rotation. Fainter clouds, however,
show high approaching velocities that are uncoupled with this apparent
disk rotation. The striking similarities between this EELR and the EELRs
around steep-spectrum radio-loud quasars provide further evidence for
the orientation-dependent unification schemes. The metal-poor gas is
almost certainly not native to the massive host galaxy. We suggest that
the close companion galaxy could be the tidally stripped bulge of a
late-type galaxy that is merging with the host galaxy. The interstellar
medium of such a galaxy is probably the source for the low-metallicity
gas in 3C\,79. 
\end{abstract}


\section{Introduction}

Low-redshift quasars are often surrounded by massive ionized
nebulae showing filamentary structures on scales of a few tens of kpcs.
More specifically, about 40\% of the quasars at $z < 0.5$ that are also
steep-spectrum (\ie\ usually FR\,II type) radio sources show such
extended emission-line regions (EELRs) with an \othree\,$\lambda5007$
luminosity greater than $4\times10^{41}$ erg s$^{-1}$. Here
we will focus our discussion on these steep-radio-spectrum quasars,
because (1) they are possibly the counterparts of FR\,II radio galaxies like 3C\,79, 
and (2) the presence of a powerful radio jet seems to be a necessary
(though not sufficient) condition for producing a luminous EELR, as
implied by the correlation between radio spectral index and extended
optical emission \citep{Bor84,Sto87}. 

An EELR typically has a mass of
$10^{9-10}$ \msun, displays globally disordered kinematics, and shows a
complex morphology that bears no obvious relationships ether with the
host galaxy or with the extended radio structure
(\citealt{Sto87,Fu06,Fu07b}; see \citealt{Sto06a} for a review). 
In terms of quasar luminosity, host galaxy luminosity, and black hole
mass, the quasars that show EELRs (hereafter the ``EELR quasars") do not
differ significantly from non-EELR quasars. But the former are all
low-metallicity quasars. The gas metallicity in their broad-line
regions (BLRs) is significantly lower ($\lesssim$ 0.6 \zsun) than that
of non-EELR quasars ($>$ \zsun) \citep{Fu07a}. Although this gas-phase
metallicity is unexpectedly low for galaxies with the masses found for
these quasar hosts, it is consistent with that in their EELRs.  
Combining all of the pieces  together, the most likely scenario for the
origin of an EELR is that it comprises gas that is native to a gas-rich
galaxy that merged with the quasar host and triggered the quasar
activity, after which a large fraction of the gas was impulsively swept
out by a large-solid-angle blast wave (\ie\ a superwind), in a manner similar to that
envisioned for quasar-mode feedback in the early universe (\eg\
\citealt{Di-05}).

3C\,79 ($z = 0.256$, 1\arcsec\ = 4 kpc)\footnote{Throughout we assume a
flat cosmological model with $H_0=70$ km s$^{-1}$ Mpc$^{-1}$,
$\Omega_m=0.3$, and $\Omega_{\Lambda}= 0.7$} is a narrow-line radio
galaxy having FR\,II radio jets with a central component
\citep{Hes95,Spa84,Har97}. 
{\it Hubble Space Telescope} (\hst) WFPC2 broad-band images show a
complex optical morphology --- a bright elliptical galaxy with an
effective radius of 7.5 kpc, accompanied by multiple ``tidal arms" and
two distinct ``cores" other than the nucleus \citep{Dun03,Kof96}. As we
will show in \S~\ref{sec:morph}, only the galaxy and the ``core"
southwest of the nucleus are true continuum sources; the rest are all
line-emitting regions. 
The nuclear spectra of 3C\,79 show stellar absorption lines and a red
continuum, indicating an old stellar population \citep{Mil81,Bor87}.  
A line-emitting region in a curving filament extending 12\arcsec\ to the
northwest of the galaxy was seen in an H$\alpha$ image and was
subsequently confirmed by a slit spectrum covering the
\othree\,$\lambda\lambda4959,5007$ region \citep{McC95,McC96}. The
spectrum also shows emission-line clouds between the filament and the
nucleus, as well as clouds extending 5\arcsec\ to the southeast.  An
\othree\,$\lambda5007$ image taken in much better seeing \citep{Sto06a}
shows a very rich morphology and an emission-line luminosity comparable
to that of the EELRs around steep-spectrum radio-loud quasars, establishing
that 3C\,79 is associated with an EELR. 

In line with the orientation-based unification schemes of FR\,II radio galaxies and
radio-loud quasars \citep{Bar89}, we believe that 3C\,79 belongs to the
same class of object as the EELR quasars, such as 3C\,249.1 and
4C\,37.43, but it is viewed at a different angle with respect to the
molecular torus surrounding the central engine. If this is true, then
the unique geometry of radio galaxies like 3C\,79 could afford us a
natural coronograph blocking the blinding glare from the central engine. Such
objects can be especially useful for studies of their host galaxies,
which potentially preserve a large amount of information regarding the
forces that have sculpted these spectacular emission-line regions. 

In this paper, we study the host galaxy and the EELR of 3C\,79 with extensive 
ground-based spectroscopy in combination with
a re-analysis of archival \hst\ WFPC2 multi-band images. First, we briefly describe our
observations and data reduction procedures in \S~\ref{sec:obs}. In
\S~\ref{sec:host} we study the morphology, stellar kinematics and
stellar populations of the host galaxy. We then demonstrate the
similarities between the 3C\,79 EELR and the ones around quasars, in
terms of gas kinematics, pressure, ionization mechanism, and metallicity
in \S~\ref{sec:EELR}. Finally, we discuss our main results in
\S~\ref{sec:dis} and close with a summary in \S~\ref{sec:sum}.

\begin{figure}[!tb]
\epsscale{1.0}
\plotone{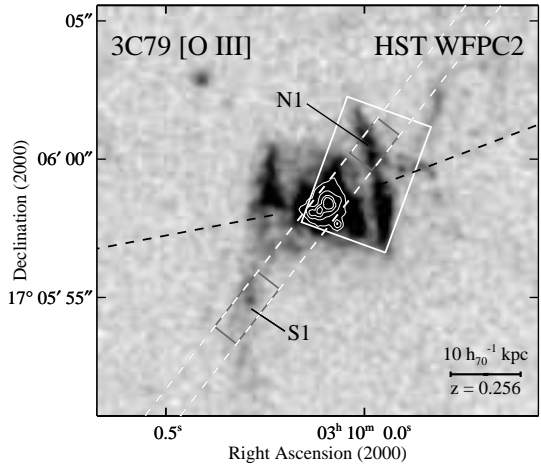}
\caption{
{\it Hubble Space Telescope} (\hst) WFPC2 FR680N \othree\ image
of 3C\,79. Superimposed are contours from the continuum dominated WFPC2
F675W image, at levels of $1.6\times10^{-17}\times$(1, 1.53, 2.34, 3.57)
erg cm$^{-2}$ s$^{-1}$ \AA$^{-1}$ arcsec$^{-2}$ above the background,
showing the structures near the nucleus. Also overlaid are the field of the GMOS/IFU ({\it white rectangular box}) and the position of the LRIS slit ({\it white dashed lines}; PA = 142$^{\circ}$). The extraction apertures for the {\it N1} and {\it S1} regions are shown with grey boxes. 
The FR\,II radio jet directions are indicated by the black dashed lines. The angular size
of the scale bar is 2\farcs51. North is up and east is to the left for
this and the following \hst\ images.  } \label{fig:obs} \end{figure}

\begin{figure*}[!tb]
\epsscale{1.0}
\plotone{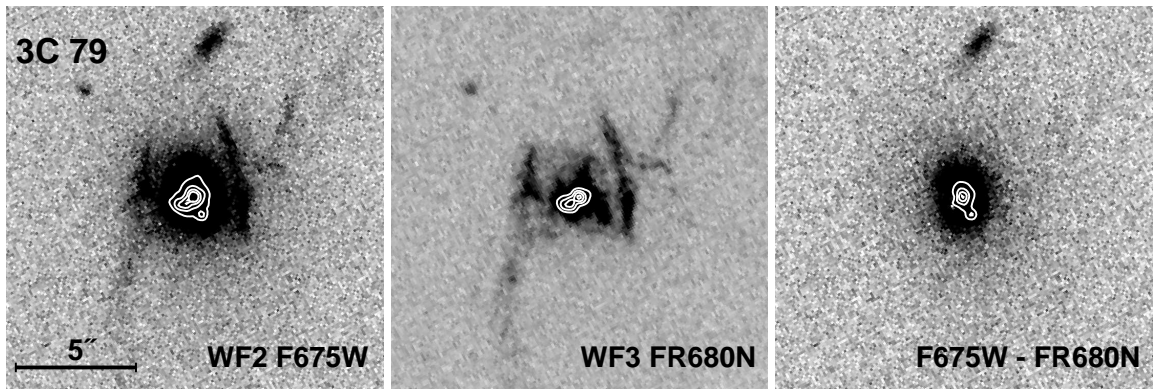}
\caption{
\hst\ WFPC2 images of 3C\,79:  ({\it left}) the F675W image, ({\it
middle}) the FR680N \othree\ image rotated, shifted and scaled to match
the F675W image, ({\it right}) the residual from subtracting the FR680N
\othree\ image off the F675W image.  The contours, which are from the
same images but smoothed by a Gaussian function with a FWHM of 3 pixels,
show the central high surface brightness features. Contour levels are
the same as in Fig.~\ref{fig:obs}.  } \label{fig:subdzl} \end{figure*} 

\begin{figure}[!tb]
\epsscale{1.1}
\plotone{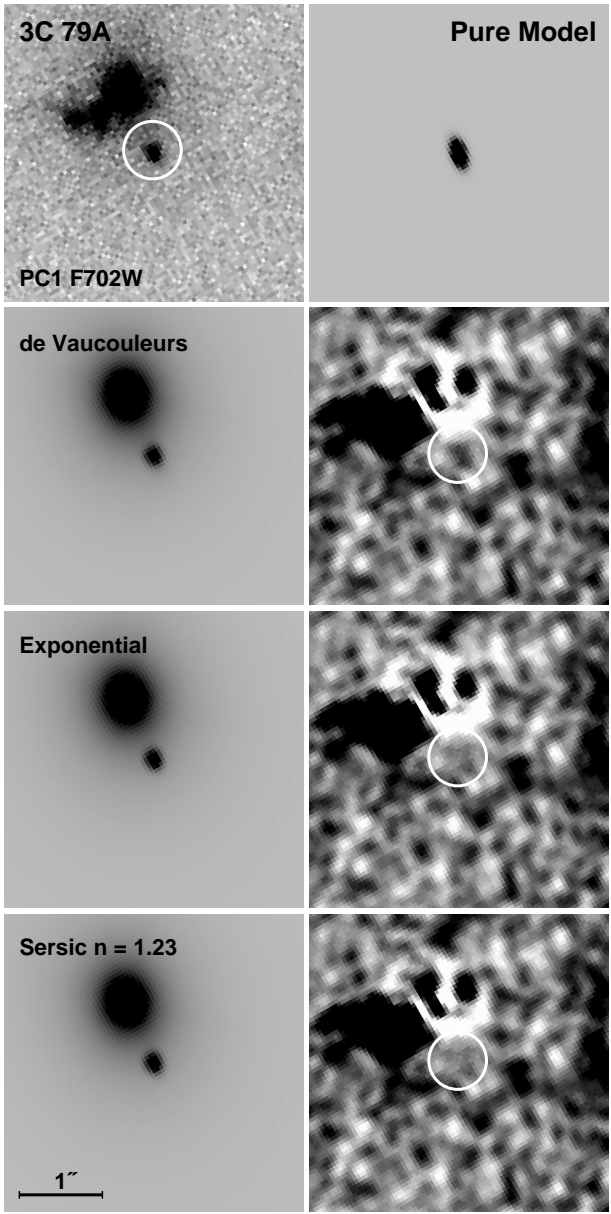}
\caption{
Two-dimentional modeling of the close companion galaxy 3C\,79A. The top
left panel shows the original WFPC2/PC1 F702W image. Directly below are
best fit de Vaucouleurs (``$r^{1/4}$ law"), exponential, and
\sersic\ models convolved with the PSF. To remove the light from the 
host galaxy and the nuclear NLR, we fitted a de Vaucouleurs plus
PSF, and the best fit models are shown together with the 3C\,79A models.
To the right of each model are shown the smoothed residuals after
subtracting the models. The residuals are displayed at a higher
contrast. The top right panel shows the best fit \sersic\ model without
convolution with the PSF, which gives the best realization of the gross
morphology of the galaxy. The white circles in some panels are centered
at the position of 3C\,79A and have a radius of 0\farcs35.  }
\label{fig:3c79a} \end{figure} 

\begin{figure*}[!tb]
\epsscale{0.75}
\plotone{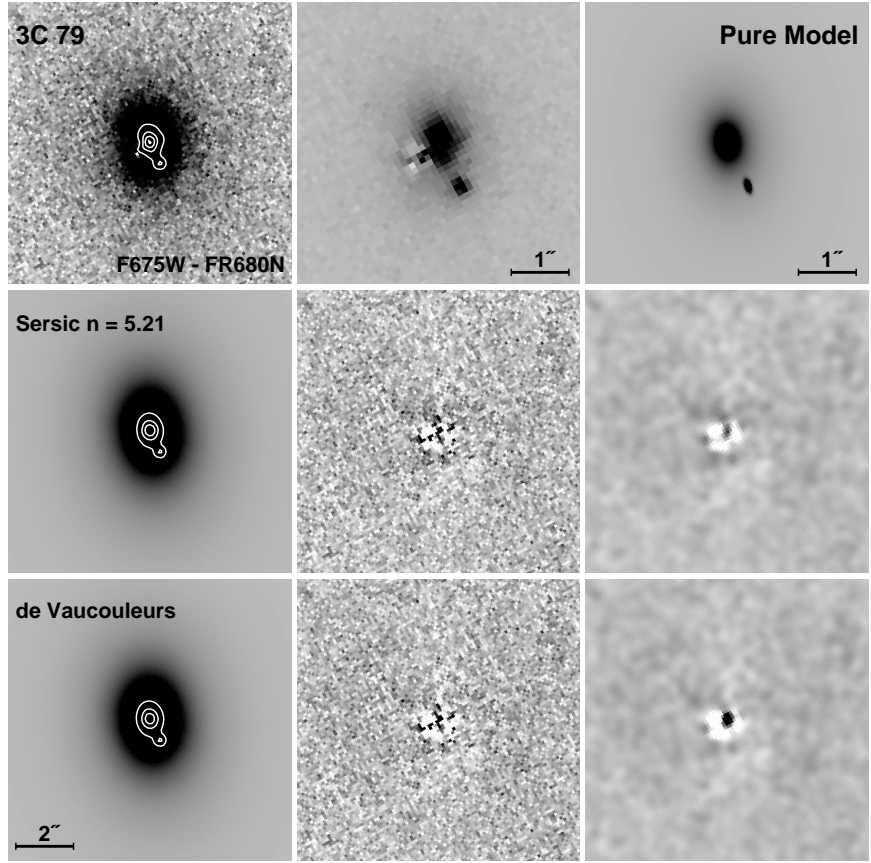}
\caption{
Two-dimentional modeling of 3C\,79 host galaxy. The top left panel shows
the WFPC2/WF emission-line free image. Directly below are best fit
\sersic\ and de Vaucouleurs models convolved with the PSF. The
superimposed contours in these panels are from the same images but
smoothed by a Gaussian function with a FWHM of 3 pixels, and are at the
same levels as those in Fig.~\ref{fig:obs}.  To the right of each model
are shown the original residual from the model subtraction and a
Gaussian smoothed version. The top middle and top right panels show
respectively the data and the best fit model (without convolution with
the PSF), both have been magnified by a factor of two and are displayed
at a lower contrast.  For the companion galaxy (3C\,79A), we used an
exponential model and have fixed its geometric parameters
($r_{1/2}$, b/a and PA) to those obtained from modeling the WFPC2/PC1
image (see Fig.~\ref{fig:3c79a}).  } \label{fig:3c79} \end{figure*} 

\begin{figure}[!tb]
\epsscale{1.2}
\plotone{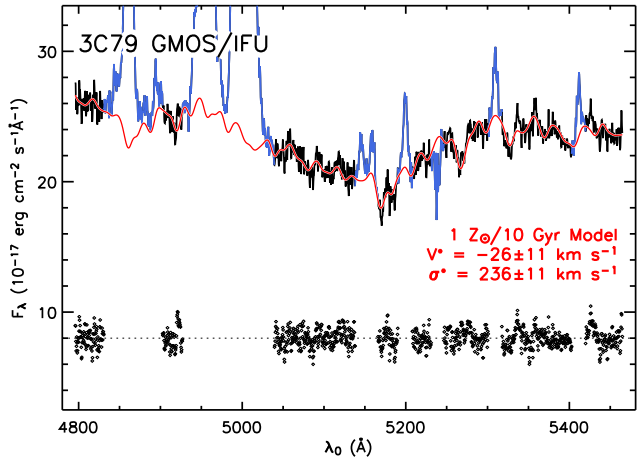}
\caption{
Modeling the kinematics of the central part ($r \lesssim 2$ kpc) of
3C\,79 host galaxy. The GMOS/IFU rest-frame spectrum of 3C\,79 host
galaxy is plotted in black, with spectral regions excluded from the
modeling highlighted in blue (including nuclear emission lines and a CCD
chip defect). Overplotted in red is the best fit instantaneous burst
model of \citet{Vaz99}, with the metallicity and age labeled in red.
Also labeled are the measured radial velocity of the stars relative to
that of the NLR ($z = 0.25632$) and the stellar velocity dispersion. The
residual after being offset by $8\times10^{-17}$ erg cm$^{-2}$ s$^{-1}$
\AA$^{-1}$ is shown with diamonds, and the dotted line going
through the residual indicates the zero level.  } \label{fig:skin}
\end{figure} 

\begin{figure*}[!b]
\hskip 0.6in
\includegraphics[angle=90,scale=0.7]{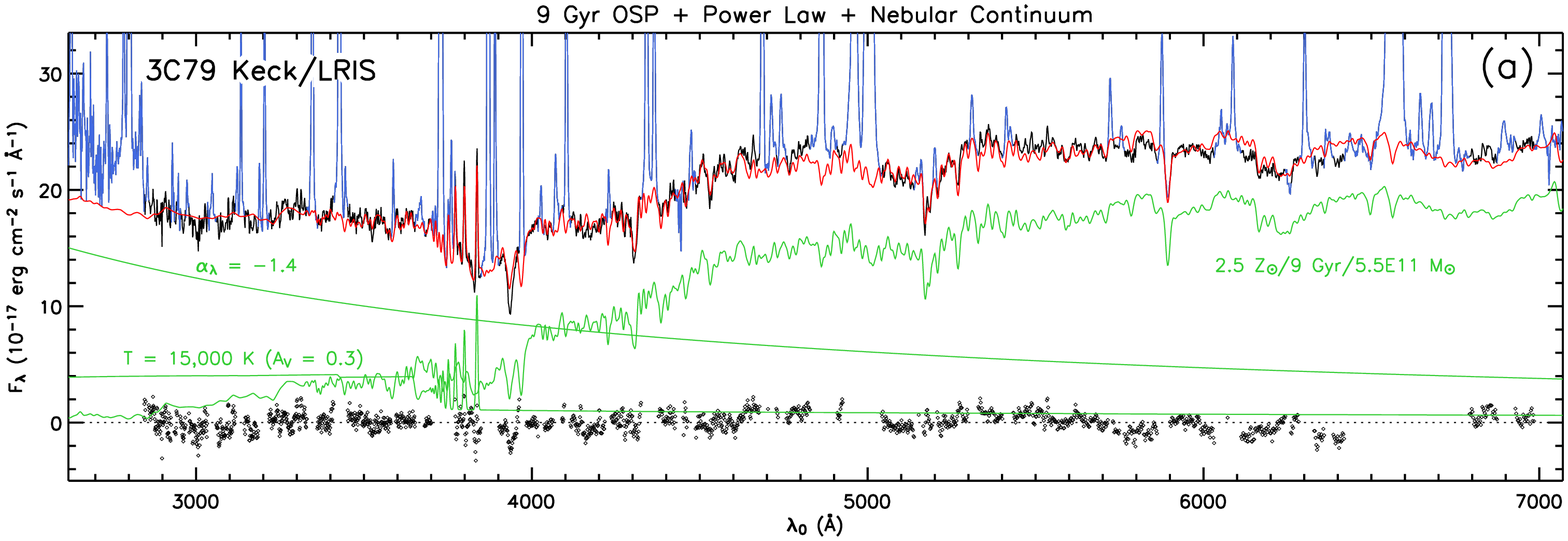}
\hskip 0.1in
\includegraphics[angle=90,scale=0.7]{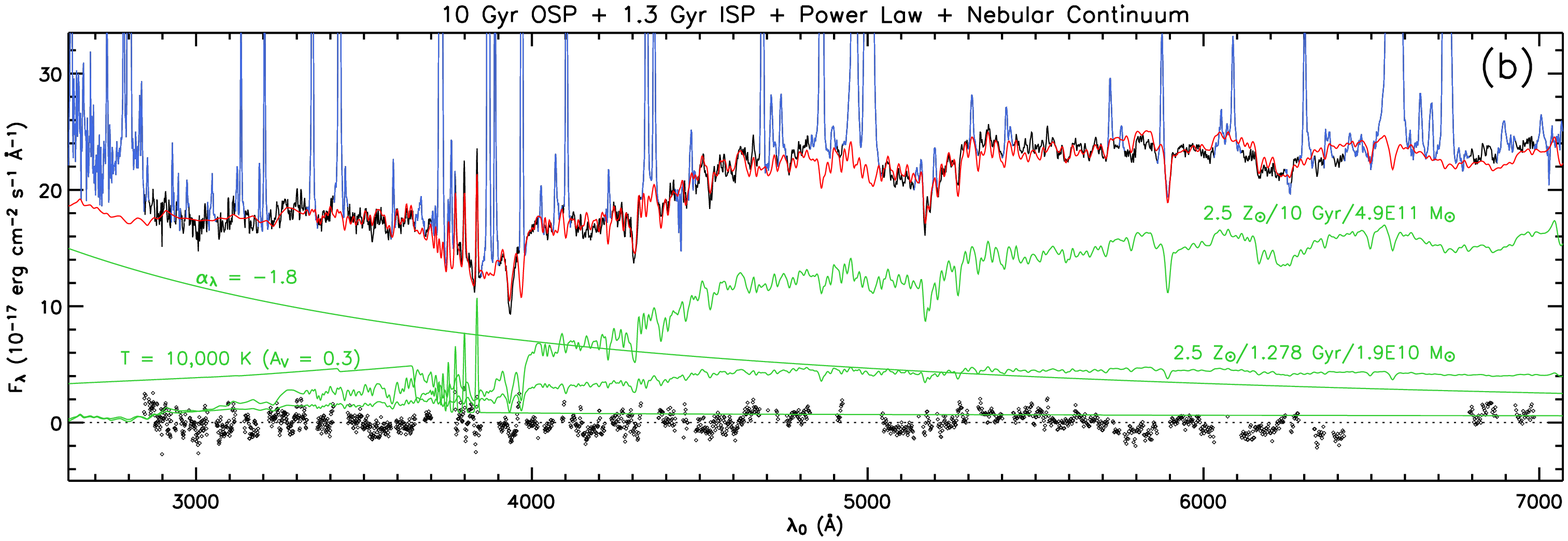}
\caption{
Decomposing the nuclear continuum of the 3C\,79 host galaxy: ({\it a}) the best fit three-component model, and ({\it b}) the best fit four-component model. 
Spectral regions that were excluded from the modeling (nuclear emission lines and imperfect sky subtraction regions) are highlighted in blue. Overplotted below ({\it green}) are \citet{Bru03} stellar population models, $\lambda^{\alpha}$ power laws representing quasar scattered light, and reddened nebular continua (including Balmer lines $\geq$ H7). The sum of the models ({\it red}) provides the best fit to the observed spectrum. Labeled are the metallicity, age and stellar mass for the stellar population models, the temperature of the nebular continuum models and the index of the power laws. The residual is shown with diamonds, and the dotted line going through the residual indicates the zero level. } 
\label{fig:spop} \end{figure*} 

\begin{figure}[!tb]
\epsscale{1.2}
\plotone{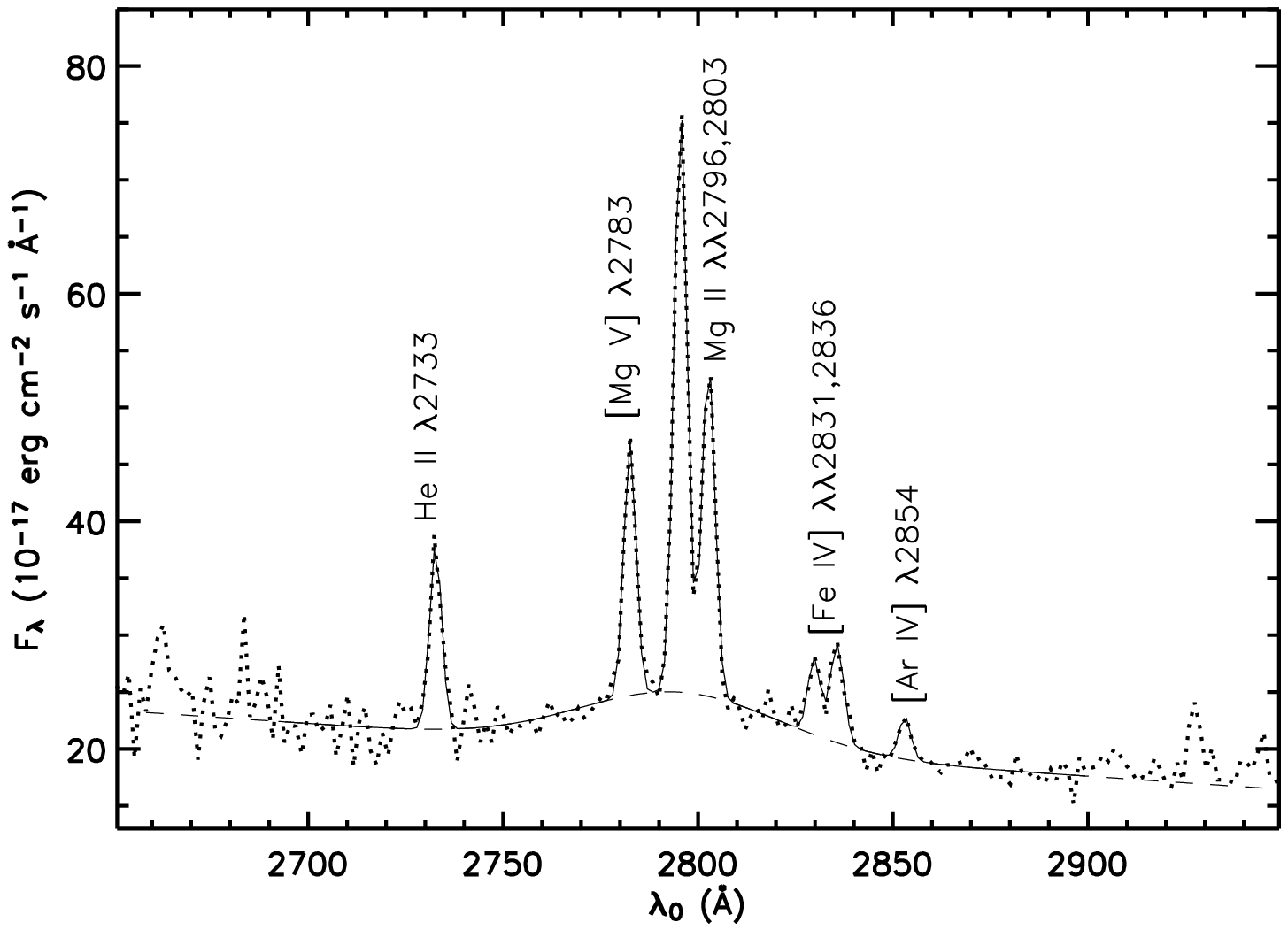}
\caption{
Detection of a broad Mg\,{\sc ii} $\lambda$2798 line in the nuclear spectrum of 3C\,79. Dotted curve shows the LRIS spectrum, the solid curve is a best fit model of eight Gaussians plus a linear continuum, and the long dashed curve shows the broad (FWHM $\sim$ 6500 \kms) Gaussian profile of the Mg\,{\sc ii} line. Key narrow emission lines are labeled.} 
\label{fig:mg2} 
\end{figure}

\section{Observations and data reduction} \label{sec:obs}
\subsection{GMOS Integral Field Spectroscopy} \label{sec:obs_gmos}

We observed 3C\,79 with the Integral Field Unit (IFU; \citealt{All02})
of the Gemini Multiobject Spectrograph (GMOS; \citealt{Hoo04}) on the
Gemini North telescope. The observations were performed on the night of
2006 December 21 (UT).  Three exposures of 2880 s were obtained using
the half-field mode with the B600/G5303 grating at a central wavelength
of 6242 \AA.  With this setting, we had a field-of-view (FOV) of
$3\farcs5\times5\arcsec$, a wavelength range of 4100 to 6900 \AA, a
dispersion of 0.46 \AA\ pixel$^{-1}$ and an instrumental full width at
half-maximum (FWHM) of 4.5 pixels (2.1 \AA). The host galaxy was placed
at the lower-left corner of the IFU field so that a large fraction of
the western part of the EELR could be covered (Figure~\ref{fig:obs}).
Feige\,34 was observed for flux calibration.

The data were reduced using the Gemini IRAF package (Version 1.8). The
data reduction pipeline (GFREDUCE) consists of the following standard
steps: bias subtraction, spectral extraction, flat-fielding, wavelength
calibration, sky subtraction, and flux calibration. Cosmic ray rejection
was performed by L.A.Cosmic \citep{Dok01} before running the data through the
reduction pipeline, with careful adjustment of the parameters to avoid
misidentification of real data.  Spectra from different exposures were
assembled and resampled to construct individual data cubes
(x,y,$\lambda$) with a pixel size of 0\farcs05 (GFCUBE). To correct for
the differential atmosphere refraction (DAR), we first binned the data
cubes along the wavelength direction to increase the S/N ratio of the
host galaxy; then the centroid of the galaxy was measured for each bin,
and a polynomial was fit to the (x,y) coordinates, restricted by the
fact that DAR only causes the centroids to vary along a straight line in
(x,y) plane (at the parallactic angle); finally we shifted the image
slices at each wavelength using the solution from the polynomial fit.
The three data cubes were then combined and binned to 0\farcs2 pixels,
the original spatial sampling of the IFU fiber-lenslet system, to form
the final data cube. 

\subsection{LRIS Long-Slit Spectroscopy} \label{sec:obs_lris}

We obtained optical long-slit spectroscopy of 3C\,79 on the night of 2007 October 17 (UT) with the Low Resolution Imaging Spectrograph \citep{Oke95} on the Keck I telescope. The Cassegrain Atmospheric Dispersion Compensator (Cass ADC\footnote{http://www2.keck.hawaii.edu/inst/adc/docs/}) was used during the observation. We took two 1200 s exposures through a 1\arcsec\ slit centered on the nucleus at a position angle of $142^{\circ}$ (Fig.~\ref{fig:obs}). On the blue arm we used the 600 groove mm$^{-1}$ grism blazed at 4000 \AA, whilst on the red arm we used the 400 groove mm$^{-1}$ grating blazed at 8500 \AA, offering, respectively, wavelength ranges of 3100$-$5600 \AA\ and 5100$-$8900 \AA\ and spectral resolutions of 4 and 7 \AA\ (FWHM). The spectra were taken at low airmass ($\sim$1.1), and the seeing was about 0\farcs7 throughout the night. 

The spectra were reduced in the standard fashion using {\sc IRAF} tasks. The blue-arm spectra were wavelength-calibrated using observations of arc lamps, whilst for the red-arm spectra we used night-sky lines. Background sky was removed by fitting low-order cubic splines to the regions on each side of the object along the slit direction using {\sc BACKGROUND}. One-dimensional spectra of 3C\,79 were first extracted with a wide (11\arcsec) aperture, and they were flux-calibrated and corrected for atmospheric absorption using the spectrophotometric standard Wolf\,1346 (taken 5 min after the nautical twilight earlier in the night). The calibrated spectra from the two arms agree perfectly in their overlapping region. Then, one-dimensional spectra of high S/N ratio were obtained by using a 2\arcsec\ extraction aperture. These spectra were multiplied by appropriate smooth curves so that their continua match those of the wide-aperture spectra. Finally, the blue and the red sides, after flux calibration and atmospheric absorption correction, were joined and binned to a common linear dispersion (1.86 \AA\ per pixel) using {\sc SCOMBINE}. The line-of-sight Galactic reddening ($A_V = 0.421$, \citealt{Sch98}) was corrected with a standard reddening curve \citep{Car89}. We obtained an absolute flux calibration by scaling the spectrum to the \hst\ photometry result from \S~\ref{sec:morph} ($F_{\lambda} = 1.8\times10^{-16}$ erg cm$^{-2}$ s$^{-1}$ \AA$^{-1}$  between 6300 and 7300 \AA).

\subsection{\hst\ WFPC2 Imaging} \label{sec:obs_hst}

To study the host galaxy morphology, we obtained WFPC2/WFC F675W
(4$\times$565 s) and FR680N (2$\times$1300 s) images, and a WFPC2/PC
F702W (2$\times$140 s) image of 3C\,79, available from the archive of
the \hst. The two broad-band images have been previously presented by
\citet{Dun03} and \citet{Kof96}. The Linear Ramp Filter (LRF; FR680N)
image has a central wavelength of 6284 \AA\ and a bandpass FWHM of 82
\AA\ at the position of 3C\,79, \ie\ the filter is centered on the
redshifted \othree\,$\lambda$5007 emission line.  Images observed with
LRFs are not flat-field calibrated in the calibration pipeline, so
we flattened the LRF image with an F631N flat field reference image. 

Additional cosmic rays after the standard pipeline reduction were
identified and replaced by L.A.Cosmic \citep{Dok01}, again with careful
adjustment of the parameters to avoid misidentification of real data.

\section{The Host Galaxy} \label{sec:host}

\subsection{Galaxy Morphology} \label{sec:morph}

Though the WFC2 F675W image is dominated by continuum radiation,
H$\beta$ and \othree\ $\lambda\lambda$4959,5007 lines fall on the
shortward wing of the filter profile, where the transmissions are about
42, 76, and 85\%, respectively. To study the morphology of the host
galaxy, one has to remove the line-emitting regions from the image. We thus rotated the
WFC3 LRF \othree\ image to the orientation of the F675W image, shifted
and scaled the image to match the F675W image, and finally subtracted it
off the wide-band image to obtain an emission-line free image (hereafter
the ``F675W$^*$'' image; Fig.~\ref{fig:subdzl}).  A close companion
0\farcs8 southwest of the major host galaxy is seen in the F675W$^*$
image, which we designate as 3C\,79A. The host galaxy itself is simply
called 3C\,79 in this section.

3C\,79A is better sampled in the PC1 F702W image, so we used this image
to study its morphology. We used C. Y. Peng's GALFIT software
\citep{Pen02} to fit the two-dimensional galaxy profile, and we used an
oversampled Tiny
Tim\footnote{http://www.stsci.edu/software/tinytim/tinytim.html} PSF.
The result is shown in Fig.~\ref{fig:3c79a}. As indicated by the best
fit \sersic\ index of $n = 1.23$, the profile is close to an exponential. 
As shown by the residuals, the exponential model clearly provides a
better fit than the de Vaucouleurs model. 

We chose to use the F675W$^*$ image to study the morphology of 3C\,79.
We converted the geometric parameters from the best fit exponential model of
3C\,79A (effective radius, axis ratio, and position angle [PA]) to the WFC image, and
froze them; however, we allowed the position and magnitude of 3C\,79A
vary in the modeling of 3C\,79. Figure~\ref{fig:3c79} shows that 3C\,79
can be described by a de Vaucouleurs model, as indicated by the high
\sersic\ index.  Table~\ref{tab:morph} summarizes the morphology
results. The PA of the minor axis of 3C\,79 is aligned
within 5$^{\circ}$ to the direction of the radio jets (as defined by the
hot spots; see Fig.~\ref{fig:obs}). The companion galaxy 3C\,79A is
located 0\farcs84 (3.3 kpc) southwest of 3C\,79, and its PA is aligned
within 7$^{\circ}$ to the direction of the host galaxy.  

\citet{Dun03} modeled the 3C\,79 host galaxy with a de Vaucouleurs model
(plus a concentric PSF to account for the narrow-line region [NLR]). For
comparison, their results are also shown in Table~\ref{tab:morph}.
\citeauthor{Dun03} used the original F675W image, \ie\ the emisson-line
regions were not removed; thus they overestimated the brightness of the
galaxy and had a poor estimate for the axis ratio and PA. We
were able to reproduce their results using GALFIT when modeling the
original F675W image.

\subsection{Stellar Kinematics} \label{sec:skin}

We extracted the host spectrum from the final combined GMOS data cube using a
$0\farcs5$ (2 kpc) radius aperture\footnote{We chose such a small aperture
because: (1) it yields the best S/N ratio compared to apertures of other
sizes; and (2) it avoids hitting the edges of the FOV at short
wavelengths}. Although the GMOS spectrum shows a poorer S/N 
ratio in comparison with the LRIS spectrum, it has a better spectral resolution (2.1 \AA\ 
FWHM as opposed of 7 \AA), making it best-suited for measuring stellar kinematics.  
We chose to use the stellar synthesis models of \citet{Vaz99} to fit the data. 
Although these models have very limited spectral
coverage compared with those of \citet{Bru03}, they have a better
spectral resolution ($\sim$ 1.8 \AA), which is similar to that of the
GMOS spectrum, and they do cover the spectral region where the spectrum show the
highest S/N ratio (S/N $\sim$ 33 at rest-frame wavelengths of 4800 \AA\
$\leq\lambda_0 \leq$ 5500 \AA). For the modeling we used M. Cappellari's
IDL program\footnote{http://www.strw.leidenuniv.nl/$\sim$mcappell/idl/}
implementing the pixel-fitting method of \citet{Cap04}. The program
finds the best fit to the data by convolving the model templates with a
line-of-sight velocity distribution (LOSVD). Since we used a Gaussian
function to parameterize the LOSVD, the program reports the mean stellar
velocity ($V^*$) and velocity dispersion ($\sigma^*$). In
Figure~\ref{fig:skin} we plot our best fit model against the data. 
Following \citet{Ems04}, we included a multiplicative
Legendre polynomial of degree 6 in the fit\footnote{Applying an additive
Legendre polynomial of similar degrees led to consistent results.} to correct 
the model continuum shape. The kinematics results are insensitive to the 
assumed model age, although $\chi^2$ increases rapidly for models younger 
than 3 Gyr and the oldest population yielded the minimum $\chi^2$. 
Using a Monte-Carlo approach, we found that the 1 $\sigma$ uncertainties
of $V^*$ and $\sigma^*$ are both about 11 \kms. These errors should be
regarded as lower limits since they do not account for the effect of the
template and continuum mismatch. Assuming a negligible amount of
systematic rotation, the measured velocity dispersion of $\sim$236 \kms\
implies a virial mass of $5 R \sigma^{2}/G = 1.3\times10^{11}$ \msun\
within 2 kpc from the galaxy center. We estimated the
luminosity-weighted velocity dispersion within $r_{1/2}$ (7.2 kpc) to be
$\sigma^*_{\rm 1/2} = 218\pm20$ \kms, with the aperture correction
function for SAURON elliptical galaxies, $(\sigma^*_{\rm
1/2}/\sigma^*_r)=(r_{\rm 1/2}/r)^{-0.066\pm0.035}$ \citep{Cap06}. The
virial mass inside $r_{1/2}$ is then $M^{\rm vir}_{\rm 1/2} =
4.0^{+0.8}_{-0.6}\times10^{11}$ \msun.

\subsection{Stellar Population} \label{sec:spop}

To study the stellar population of the host galaxy, we have performed detailed continuum modeling to the 3C\,79 nuclear spectrum obtained from LRIS.
The host galaxy shows a strong UV excess (Fig.~\ref{fig:spop}) and a broad (FWHM $\sim$ 6500 \kms) Mg\,{\sc ii} $\lambda$2798 line (Fig.~\ref{fig:mg2}), but no broad H$\alpha$ or H$\beta$ lines are seen. These characteristics seem to resemble closely the radio galaxy Cygnus A \citep{Ant94,Ogl97}, implying a significant scattered quasar component. In addition, stellar absorption features are evident (Ca\,{\sc ii} K $\lambda$3933, the G band $\lambda$4300, and the Mg\,{\sc i} {\it b} $\lambda$5175, etc.), implying a dominating old stellar population (OSP). As a first guess, we tried to model the continuum with three components: (1) a nebular continuum (at $T$ = 10,000, 15,000, and 20,000 K), (2) a power law ($F_{\lambda} \propto \lambda^{\alpha_{\lambda}}; -0.4 \leq \alpha_{\lambda} \leq -1.8$ in accordance with the composite UV/optical quasar spectrum; \citealt{Van01}), and (3) an OSP (7 Gyr $\leq$ Age $\leq$ 10 Gyr\footnote{At $z$ = 0.256 the age of the Universe is 10.4 Gyr.}, and $Z$ = 0.2, 0.4, 1.0, and 2.5 \zsun; \zsun\ = 0.02). 
The same IDL program was used in the modeling as in the previous section, 
but this time we did \emph{not} include polynomials to adjust the model continuum shape. 
Since the nebular continuum scales with the Balmer lines, we required that the H$\gamma$ flux from the models to be within 5\% of the measured flux.     

Figure~\ref{fig:spop}{\it a} shows the best fit three-component model (\ie\ the model that produces the least $\chi^2$), which is a combination of a 9-Gyr-old 2.5 \zsun\ population, a nebular continuum at $T$ = 15,000 K, and a $\lambda^{\alpha}$ power law with $\alpha = -1.4$. This model fits both the continuum shape and most of the absorption features, however, it underpredicts the Ca\,{\sc ii} K line and the G band.
This inadequacy can be removed once we introduce a fourth component --- an intermediate-age stellar population (1 Gyr $\leq$ Age $\leq$ 3 Gyr, and $Z$ = 0.2, 0.4, 1.0, and 2.5 \zsun). Figure~\ref{fig:spop}{\it b} displays the best fit four-component model, which successfully reproduces the Ca\,{\sc ii} K line and the G band absorption.
The global best fit to the data is given by stellar population
models with a metallicity of 2.5 \zsun. In order to gauge how robust
this metallicity result is, we compared the best fit four-component
models of different metallicities. The best fit 1.0 \zsun\ model
gives a generally good fit to both the absorption features and the
overall continuum shape, except that it cannot reproduce the high S/N
continuum shape around the Mg\,{\sc i} {\it b} feature 
between 5000 and 5500 \AA\
(rest-frame), resulting in a 16\% increase in the $\chi^2$ value
relative to the 2.5 \zsun\ model. Models with sub-solar metallicities
all fail to give an adequate fit, especially in the continuum below
4300 \AA\ (rest-frame). We thus conclude that the stellar population
in the host galaxy has a super-solar metallicity. A conservative
range of metallcities that can give an adequate fit to the LRIS
spectrum is between 1 and 2.5 \zsun.

We have also attempted to use a young stellar population ($<$ 0.1 Gyr) to replace the power law component, but the best fit model underpredicts the higher ($\geq$ H7) Balmer lines if it correctly predicts the H$\gamma$ flux (due to the Balmer absorption lines from the young population), and it gives a much poorer fit to the rest-frame UV continuum (2900\AA\ $< \lambda_0 <$ 3800\AA), in particular, it overpredicts the Balmer limit break at 3650 \AA.

If we adopt the average total stellar mass from the two best fit models 
shown in Fig.~\ref{fig:spop}, then the stellar
mass inside $r_{\rm 1/2}$ is $M^*_{\rm 1/2} =
2.6\times10^{11}$ \msun. The $M^*_{\rm 1/2}/M^{\rm vir}_{\rm 1/2}$ ratio implies a dark matter fraction of $\sim34$\% within one
effective radius, in agreement with the dark matter fraction of nearby
early-type galaxies \citep{Cap06}. 
We caution that there are significant uncertainties in both the
stellar mass derived from single-burst stellar synthesis models and
the dynamical mass using the virial theorem. Hence, this
straightforward calculation of dark matter content is subject to an
error of at least a factor of two.
We note that the absolute $R$-band
magnitude\footnote{Converted from the F675W ST system to $R$-band Vega
system ($-$0.7 mag), after applying $k$-correction ($-$0.27) and
passive evolution correction (+0.23), following \citet{Lab06}.} of the
galaxy is about $-23.2$, which is the same as that of the host galaxies
of EELR quasars \citep{Fu07a}.

The compact companion galaxy 3C\,79A is detected in the GMOS data cube. The comparison of the spectrum extracted from 3C\,79A and a background spectrum from a symmetric location with respect to the minor axis of the host galaxy shows a red continuum\footnote{The continuum shape did not change much after we corrected it for the ``fixed-aperture effect" \citep{Fu07b}.}, indicating an old stellar population for 3C\,79A. This galaxy is reminiscent of the close companion galaxy 1\arcsec\ away from the EELR quasar 3C\,48 \citep{Sto07}, which is also elongated towards the AGN nucleus. The S/N ratio of the spectrum does not allow a detailed modeling. 

\begin{figure*}[!tb]
\epsscale{1.0}
\plotone{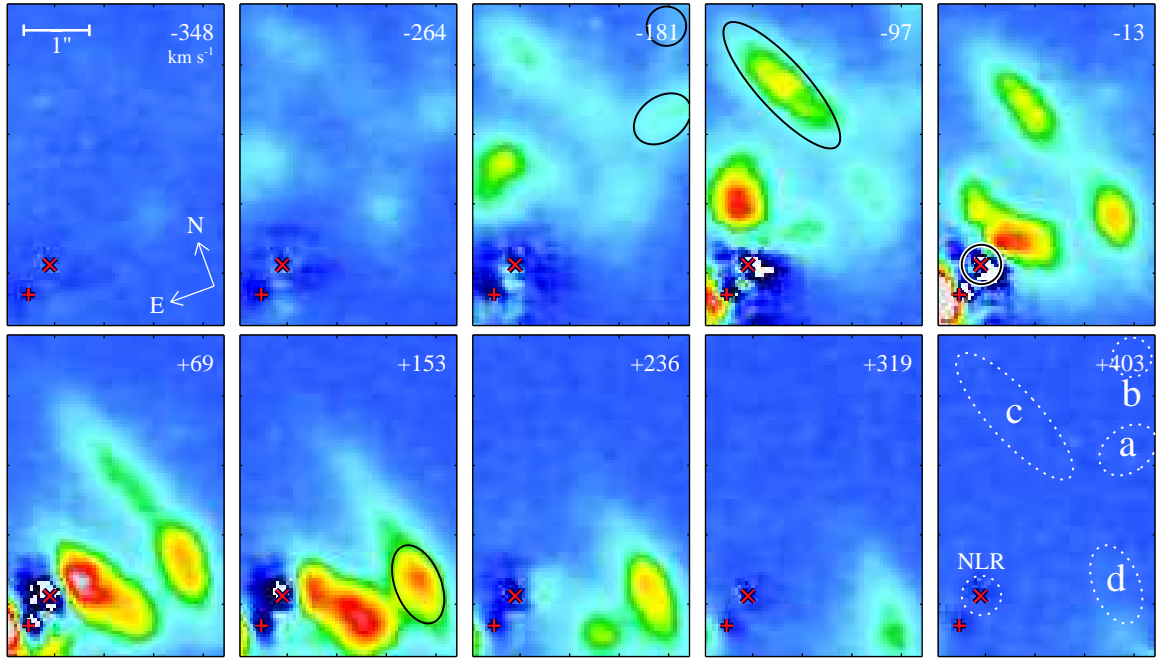}
\caption{
GMOS/IFU \othree\,$\lambda$5007 radial velocity channel maps for the
3C\,79 EELR. The central velocities, in \kms, relative to that of the
quasar NLR, are indicated in the upper right corner of each panel. The
red multiplication signs mark the nuclear narrow-line region, and the
red plus signs indicate the brightest nearby emission-line cloud, both
of which have been subtracted from the data cube to show the EELR more
clearly. Solid ellipses show the extraction apertures of the five
emission-line clouds discussed in \S~\ref{sec:EELRspec} in individual
panels corresponding to their central velocities. In the last panel
these apertures are shown all together as dashed ellipses, along with
their names labeled nearby.  } \label{fig:cmap} \end{figure*} 

\begin{figure*}[!t]
\epsscale{1.1}
\plotone{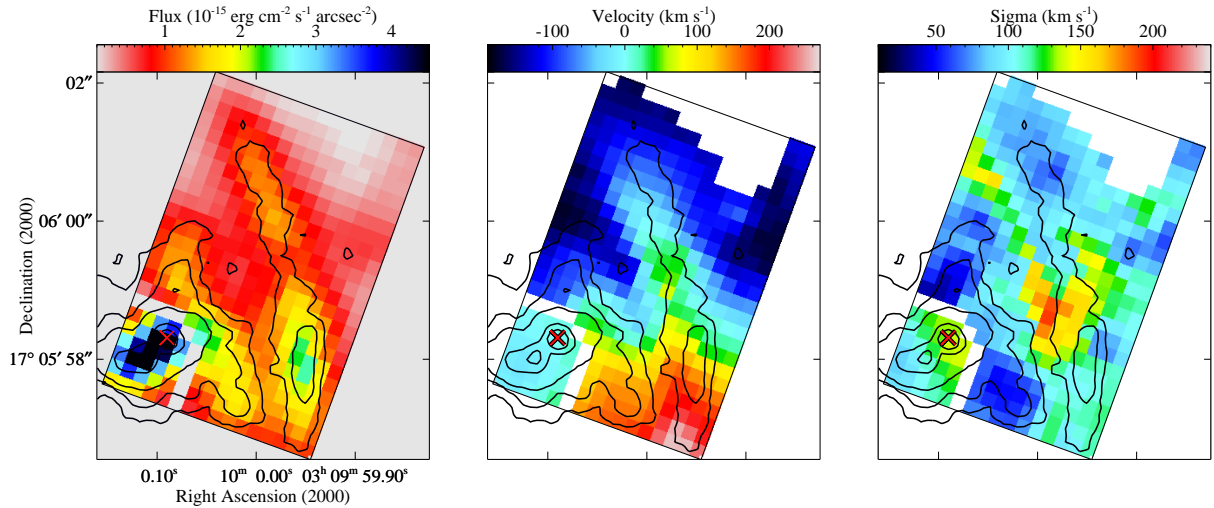}
\caption{
GMOS/IFU \othree\,$\lambda$5007 velocity field of the 3C\,79 EELR.
Inside the FOV there is a white no-data boundary near the nuclear
region. The velocity field outside of the boundary is measured from a
data cube where both the nuclear NLR and the bright emission-line cloud
directly to the east were removed (see also Fig.~\ref{fig:cmap}), while
inside the boundary the data come from the original data cube. The
fluxes inside the boundary were scaled down by a factor of 6 to show the
high surface brightness peaks.  Overlaid are contours from the \hst\ WF3
LRF \othree\ image smoothed by a Gaussian function with a FWHM of 1.5
pixels; contours are at levels of $1.8\times10^{-15}\times(1 ,2 , 4, 20,
40)$ erg cm$^{-2}$ s$^{-1}$ arcsec$^{-2}$.  The three panels, from left
to right, are line intensity, radial velocity (relative to that of the
nuclear NLR) and velocity dispersion maps. Pixels are 0\farcs2 squares.
The red multiplication signs indicate the center of the host galaxy.  }
\label{fig:vfield} \end{figure*} 

\begin{figure*}[!tb]
\epsscale{0.58}
\plotone{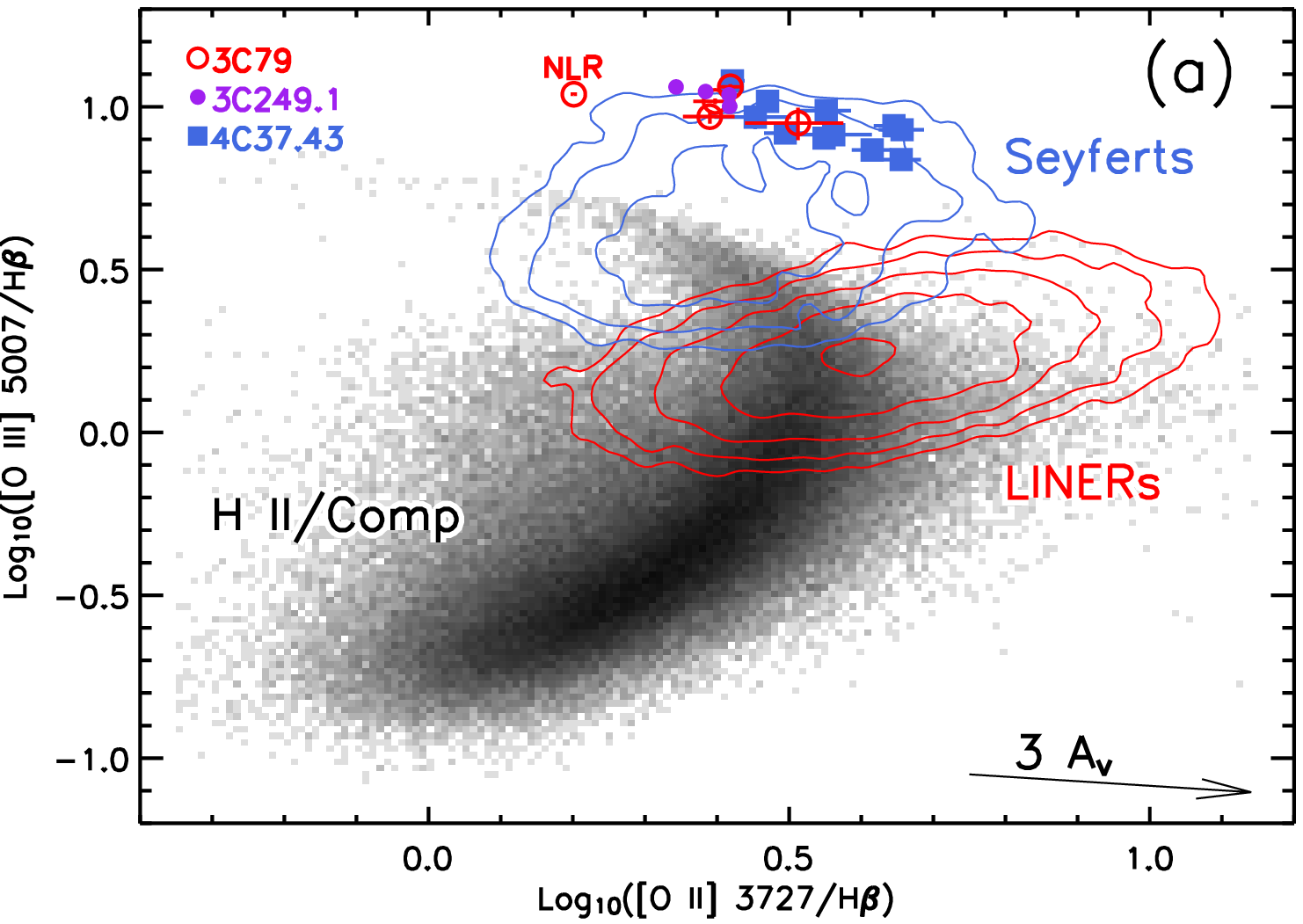}
\plotone{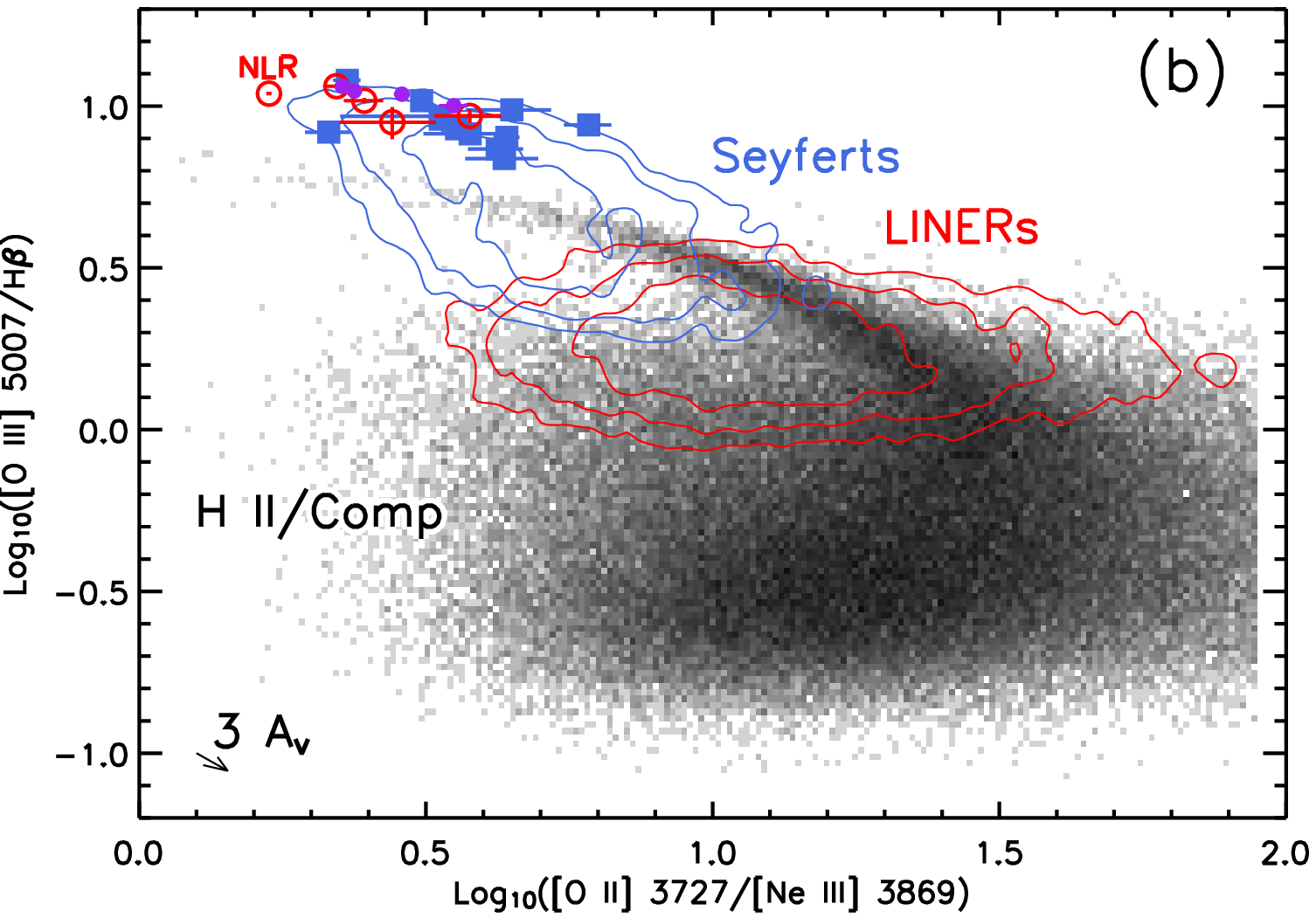}
\caption{
Log-scaled density distributions of {\it SDSS} Seyferts (blue contours),
LINERs (red contours), and star-forming galaxies (including
starburst-AGN composite galaxies; the background image) in line-ratio
diagrams: ({\it a}) \otwo\,$\lambda\lambda3726,3729$/H$\beta$ vs.
\othree\,$\lambda5007$/H$\beta$, and ({\it b})
\otwo\,$\lambda\lambda3726,3729$/[Ne\,{\sc iii}]\,$\lambda3869$\ vs.
\othree\,$\lambda5007$/H$\beta$.  Measurements from the EELRs of 3C\,79,
3C\,249.1, and 4C\,37.43 are shown as red open circles, violet solid
circles and blue squares, respectively \citep{Fu06,Fu07b}.  Both
diagrams clearly show that the EELRs are photoionized by an AGN-type
ionizing spectrum.  } \label{fig:class} \end{figure*} 

\section{The Extended Emission-Line Region} \label{sec:EELR}

The \hst\ LRF \othree\ image (Fig.~\ref{fig:obs}) shows not only the filament seen
in the H$\alpha$ image of \citet{McC95}, but also luminous emission-line
``arms'' at distances around 2\farcs5 from the nucleus, and an
``inner arc'' 1\arcsec\ to the west. To compare the
EELR of 3C\,79 with those around quasars, we determined the
total luminosity in the \othree\,$\lambda5007$ line within an annulus of
inner radius 11.2 kpc and outer radius 44.9 kpc\footnote{This is
equivalent to the 10 kpc/40 kpc annulus used by \citet{Sto87} since they
assumed an empty universe with $H_0=75$ km s$^{-1}$ Mpc$^{-1}$ and $q_0
= 0$.} (\lothree). We found a luminosity of \lothree\ = $7.8\times10^{42}$ erg s$^{-1}$, close to that of the most luminous quasar EELR at $z < 0.5$, \ie\
that of 4C\,37.43, \lothree\ = $9.4\times10^{42}$ erg s$^{-1}$
(converted from \citealt{Sto87}). As a reference, the least luminous
detected EELR in the radio-loud quasar subsample of \citet{Sto87} has a
luminosity about 10 times lower (PKS\,2135$-$147, \lothree\ = $6.9\times10^{41}$ erg
s$^{-1}$).

\subsection{Gas Kinematics} \label{sec:gaskin}

Kinematics of the ionized gas can be measured from strong emission
lines. Since a single 48 min GMOS exposure is enough to acquire a good S/N
ratio in the \othree\,$\lambda$5007 line region, we derived the velocity
fields from the single data cube that has the smallest airmass (AM =
1.172) and the best seeing.  To study the velocity structure for the
clouds in the inner arc that is only 1\arcsec\ from the nuclear NLR, we
subtracted the light from the NLR and the bright nearby cloud 0\farcs5
east to the NLR by simultaneously fitting two Moffat PSF profiles at
each wavelength in the wavelength range where both clouds are
apparent\footnote{The two clouds have essentially the same velocity, at
$z = 0.25632\pm0.00001$.}. After trials it was found that the best
subtraction results came from using the best fit Moffat profile from
modeling the Feige\,34 data cube over the same wavelength range,
indicating both data cubes have the same resolution (FWHM $\simeq$
0\farcs6 at the \othree\ region). We then fixed the Moffat parameters 
to those of the standard star and let the positions and
magnitudes vary. Finally, we fixed the positions to the average values
across the wavelength range from the previous fit and performed the fit
with only the magnitudes as free parameters. The resulting
PSF-subtracted data cube is presented as a set of velocity channel maps
in Fig.~\ref{fig:cmap}. The velocity field is shown in
Fig.~\ref{fig:vfield}. The velocity dispersion measurements
have been corrected for the $\sigma_0 = 42$ \kms\ instrumental
resolution. Due to the imperfect PSF subtraction process, no useful
kinematics can be extracted from the PSF-subtracted data cube in the
nuclear region bordered by the white no-data pixels in
Fig.~\ref{fig:vfield}. Hence, we measured the velocity field in this
region from the original data cube and inserted the results into the
final velocity field. 

Most of the extended emission comes from two main filaments --- the
inner arc and the outer arm 2\farcs5 from the nucleus. Both filaments
show similar velocity gradients, with the northern part approaching
towards us and the southern part receding from us at velocities of
$\sim200$ \kms. Together, these two filaments are consistent with a
common disk rotation with a flat rotation curve. This sort of locally
ordered kinematics of the brighter clouds is reminiscent of that of the
three most luminous clouds southeast of 3C\,249.1 \citep{Fu06}. However,
the much fainter clouds outside of the outer arc disrupt this simple
picture by showing large approaching velocities.  

The velocity dispersion over most of the region is around 50 to 120 \kms.  The region
showing large $\sigma$ ($150-200$ \kms) between the inner arc and the
outer arm, although roughly coincident with the jet direction, in fact also has
multiple velocity components that happen to contribute light to the same
pixels.    

Overall, the velocity field of the 3C\,79 EELR is similar to those of the
quasar EELRs, in the sense that all of them appear globally
disordered but locally ordered, and they all show super-sonic velocity
dispersions (sound speed $c_s \sim 17$ \kms) within the same range
between 50 and 130 \kms\ \citep{Fu06,Fu07b}. 

\subsection{Spectra of Emission-Line Clouds} \label{sec:EELRspec}

To increase the S/N ratio, we have combined GMOS spectra within various
regions. The spectra of the clouds in the inner arc ($r \leq 1\farcs2$)
are heavily contaminated by light from the nuclear NLR, making it
difficult to measure their line intensities accurately.  We therefore
concentrated on the EELR clouds in the outer arm and beyond. In
Fig.~\ref{fig:cmap} we have shown the extraction apertures for the four
EELR clouds.  We have also extracted a NLR spectrum using a circular
aperture with a radius of 0\farcs3. In order to measure the key metallicity 
diagnostic --- the \ntwo\,$\lambda6584$/H$\alpha$ ratio --- we extracted 
LRIS spectra from three regions along the slit: a 2\arcsec-wide
aperture centered on the nucleus was used for the nuclear NLR, and the 
apertures used for the {\it N1} and {\it S1} regions are shown in Fig.~\ref{fig:obs}.   

Unlike those for the EELR clouds, NLR spectra should be corrected for 
stellar absorption features. We fitted the continuum using the same stellar synthesis model templates as
in \S~\ref{sec:spop} and subtracted off the best fit model to obtain
the clean emission-line spectrum of the NLR.  Note that the line ratios
for the NLR should be reliable even if the continuum subtraction were
just barely correct, given the large equivalent widths of the emission
lines that are affected by strong absorption features.  Intrinsic
reddening due to dust in front of {\it or} associated with the cloud was
determined from the measured H$\gamma$/H$\beta$ ratio for GMOS spectra and from H$\alpha$/H$\beta$ ratio for LRIS spectra, assuming theoretical ratios of 0.468 for H$\gamma$/H$\beta$ (the value for case B recombination at $T =
10^4$ K and $N_e = 10^2$ \cc; \citealt{Ost89}) and 3.1 for H$\alpha$/H$\beta$. 
For clouds $a$ and $b$, where the S/N ratio of the H$\gamma$ line is poor, we assumed zero
intrinsic reddening. For clouds $N1$ and $S1$, where the measured H$\alpha$/H$\beta$ is slightly lower than 
the theoretical value, we also assumed zero intrinsic reddening.
Both the intrinsic reddening and Galactic
reddening were corrected using a standard Galactic reddening law
\citep{Car89}.  

The reddening-free emission-line fluxes and 3 $\sigma$
upper limits are tabulated in Table~\ref{tab:flux}. The quoted 1
$\sigma$ errors were derived from the covariance matrices associated
with the Gaussian model fits, with the noise level estimated from
line-free regions on either side of an emission line. 

There is enough overlapping area to show that the data from the two instruments agree fairly well:
(1) The low intrinsic reddening for cloud $c$ is consistent with the measured H$\alpha$/H$\beta$ ratio of 3.09 in region $N1$;
(2) The GMOS NLR spectrum suggests an intrinsic reddening of $A_V$ = 0.28, whilst the LRIS NLR spectrum shows $A_V$ = 0.36;
(3) The individual line ratios relative to H$\beta$ of regions $c$ and NLR (GMOS) are very much consistent with those measured from the LRIS $N1$ and NLR spectra, respectively.  

\subsubsection{Electron Density and Temperature} \label{sec:ne_t}

The \otwo\ luminosity-weighted average electron density ($N_e$) and
electron temperature ($T_e$) can be measured from the intensity ratio of
the \otwo\,$\lambda\lambda3726,3729$ doublet and the \othree\
($\lambda4959+\lambda5007$)/$\lambda4363$ ratio, respectively
\citep{Ost89}. Our GMOS data cube covers all of these five lines, allowing us
to measure the density and temperature in individual emission-line
clouds. The \stwo\,$\lambda\lambda6717,6731$ doublet is covered by the
LRIS spectra, however, they fall into the region where strong OH night sky lines
dominate and severe CCD fringing occurs, preventing reliable measurements. 
We thus only discuss results from the \otwo\ doublet.

The \otwo\ doublet is barely resolved in our spectra. However, for the
NLR and the EELR clouds $c$ \& $d$, the high S/N ratio profile of the
doublet allows a meaningful decomposition once we constrain their
expected wavelengths to be at the same redshift as that of the nearby
[Ne\,{\sc iii}]\,$\lambda3869$ line. 
Uncertainties of the line fluxes from this fit 
tend to be low as the wavelengths are no longer 
free parameters. Lifting this wavelength constraint gives a
more conservative error estimate. Here we have adopted the mean values of the error estimates from the two fits --- one with and one without wavelength constraints.
The weak \othree\,$\lambda4363$
line is detected in all the clouds except $b$.  We then derived $T_e$
and $N_e$ consistently using the IRAF routine TEMDEN for the three
regions where both \otwo\ and \othree\ ratios are available.
Table~\ref{tab:prop} summarizes the results. For the clouds in the EELR,
the temperatures are about 12,000 to 14,000 K, and the densities are
about 100$-$200 \cc, similar to those of quasar EELRs that were measured
using the same technique \citep{Sto02,Fu06,Fu07b}. 

\subsubsection{Ionization Mechanisms}

The emission-line ratios of the 3C\,79 clouds are very similar to those
of the quasar EELRs. Previously we have concluded that the spectra of
quasar EELRs are most consistent with being photoionized by quasar
continuum, and are inconsistent with shock or the self-ionizing ``shock
+ precursor" models \citep{Fu07b}. Now we re-emphasize this point by
comparing strong line ratios of the EELRs of the radio galaxy 3C\,79 and
the quasars 3C\,249.1 and 4C\,37.43 with emission-line galaxies at a
similar redshift range.  The Sloan Digital Sky Survey Data Release 4
(SDSS DR4) provides high-quality optical spectra for a large number of
galaxies. The emission-line fluxes after correcting for stellar
absorption and foreground extinction are publicly
available\footnote{http://www.mpa-garching.mpg.de/SDSS/DR4/; and see
\citet{Tre04} for a description of the data.}. We first created a
subsample of SDSS emission-line galaxies by including only those with
S/N $>$ 3 in the strong emission-lines \otwo\,$\lambda3727$ (=$\lambda3726+\lambda3729$), H$\beta$, \othree\,$\lambda5007$, H$\alpha$, \ntwo\,$\lambda6584$ and
[S\,{\sc ii}]\,$\lambda\lambda6717,6731$. Then we divided the sample
into star-forming galaxies and AGN using the empirical dividing line of
\citet{Kau03a}, and corrected for the intrinsic reddening using the Balmer
decrement and the standard Galactic reddening curve \citep{Car89}. The
theoretical values for the H$\alpha$/H$\beta$ ratio are 2.85 for
star-forming galaxies and 3.1 for AGN. Finally, we classified the
galaxies into Seyferts, LINERs, star-forming galaxies and
AGN/starforming composites, following the classification schemes of
\citet{Kew06}. 

Figure~\ref{fig:class} compares the locations of our EELRs and the
distributions of the SDSS galaxies in two line-ratio diagnostic
diagrams. As can be seen, the EELRs share the same territory as the
Seyfert galaxies, but are clearly distinguishable from star-forming and
composite galaxies, indicating that the EELRs are photoionized by an
AGN-type ionizing spectrum. 

The line-ratios from the 3C\,79 EELR match those from 3C\,249.1 and
4C\,37.43 very well, thus pure shock and ``shock + precursor" models can
be ruled out, since it has been shown (using the same line-ratios) that these models do not fit
these quasar EELRs (see Fig. 5 of \citealt{Fu07b} for an example).  

\begin{figure*}[!tb]
\epsscale{0.58}
\plotone{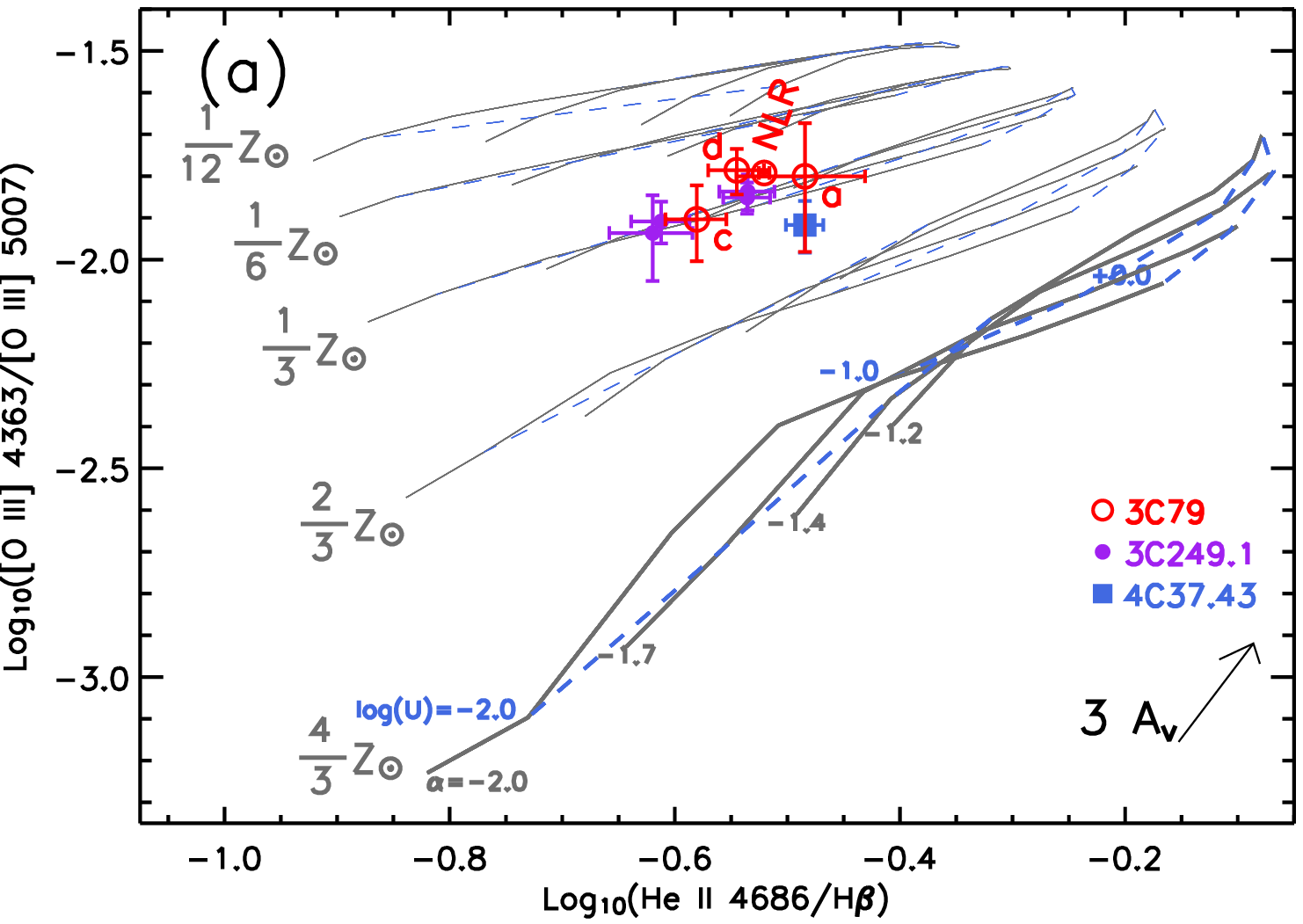}
\plotone{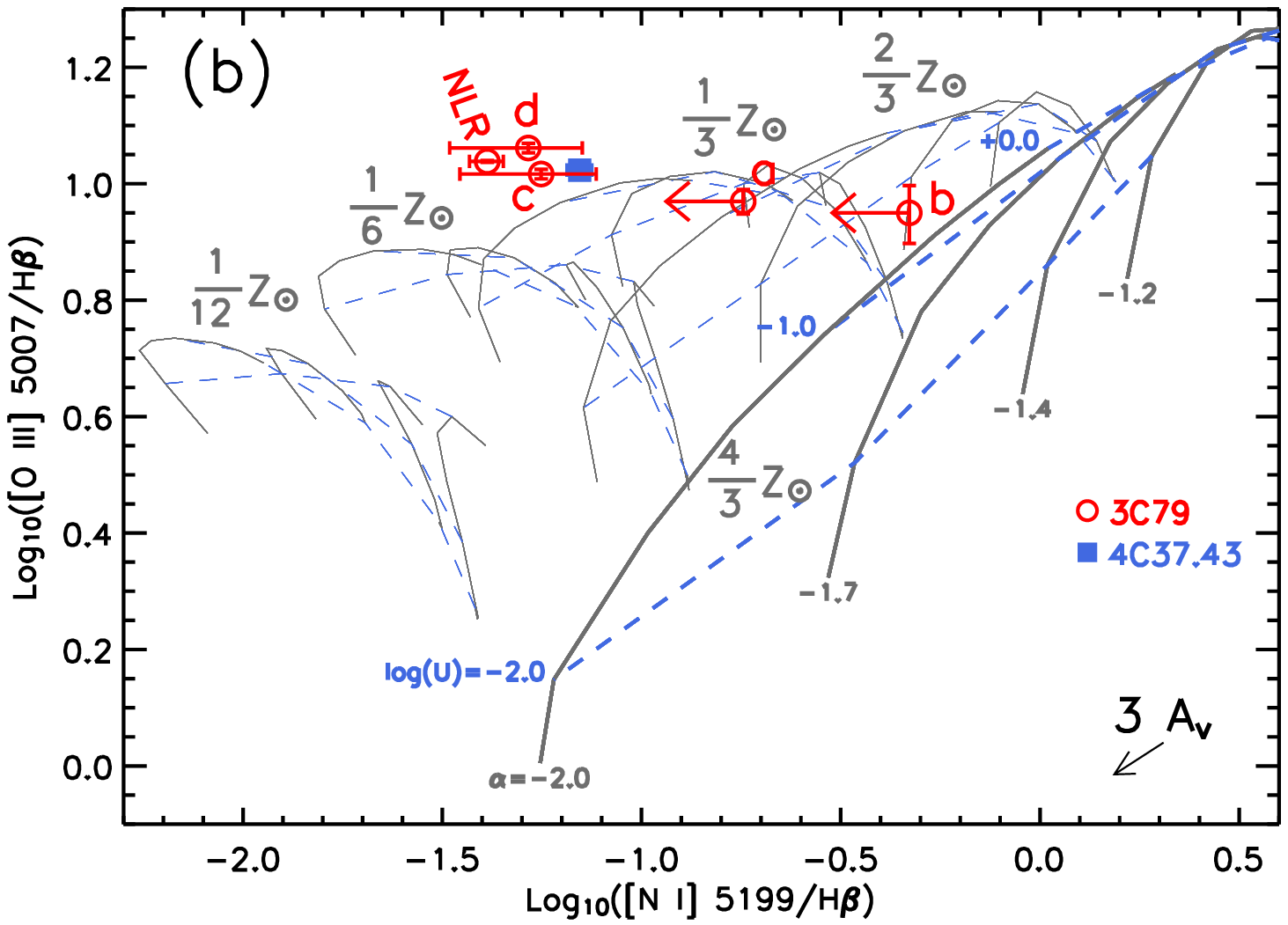}
\caption{
Metallicity-sensitive line-ratio diagrams: ({\it a})
\hetwo\,$\lambda4686$/H$\beta$ vs.
\othree\,$\lambda4363$/\othree\,$\lambda5007$, and ({\it b})
\none\,$\lambda5199$/H$\beta$ vs. \othree\,$\lambda5007$/H$\beta$.
Measurements from the EELRs of 3C\,79, 3C\,249.1, and 4C\,37.43 are
shown as red open circles, violet solid circles and blue squares,
respectively \citep{Fu06,Fu07b}. Besides 3C\,79, accurate \none\ flux
measurements are only available for 4C\,37.43 E1 \citep{Sto02}.
Photoionization model grids from \citet{Gro04a} are shown for five
metallicities from 0.1 to 1.3 \zsun. For each metallicity, model
predications are given for a range of ionization parameters (-2.3 $\leq$
log($U$) $\leq$ 0) and four power law indices representing the quasar ionizing continuum ($F_{\nu}
\propto \nu^{\alpha_{\nu}}$; $\alpha_{\nu} = -1.2, -1.4,
-1.7$, and $-2.0$).  The grids for the highest metallicity are
highlighted to show the inconsistency between our data and predictions
assuming super-Solar metallicity gas.  Both diagrams show that the EELR
of 3C\,79 consists of sub-solar metallicity gas, matching the gas
metallicity of the EELRs of 3C\,249.1 and 4C\,37.43 \citep{Fu07b}, as
well as the metallicity of the BLRs of radio-loud quasars showing EELRs
\citep{Fu07a}.  } \label{fig:z} \end{figure*} 

\begin{figure}[!tb]
\epsscale{1.2}
\plotone{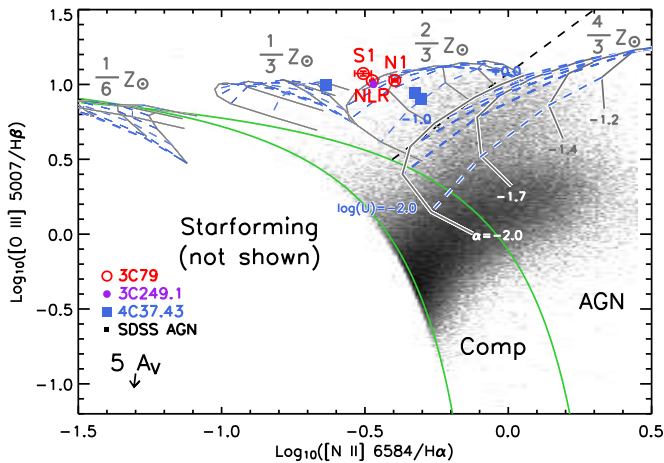}
\caption{
The nuclear NLR and EELR of 3C\,79, along with quasar EELRs, are much metal-poorer than in typical AGN, 
as seen in the metallicity diagnostic diagram of \ntwo\,$\lambda6584$/H$\alpha$ vs.
\othree\,$\lambda5007$/H$\beta$. Keys are the same as in Fig.~\ref{fig:z}.
The background image shows the density distribution of the SDSS AGN sample in log scale. Objects above the upper green curve are objects dominated by AGN (LINERs are concentrated in the lower denser branch and Seyferts in the upper branch; \citealt{Kew06}), below the lower curve are starforming galaxies \citep{Kau03a}, and AGN/starforming composite galaxies are in between. Most Seyferts can be fit quite well with the super-solar metallicity models. The top region bordered by the black dashed line and the upper green curve is where the low-metallicity Seyferts are located \citep{Gro06}.} \label{fig:n2} \end{figure} 

\subsubsection{Gas Metallicity}

With the knowledge that the EELR of 3C\,79 is photoionized by the hidden
quasar, we can now estimate the metallicity of the gas using
photoionization models. In the following we will compare our data with
the dusty radiation-pressure dominated photoionization models of
\citet{Gro04a} with a density of 1000 \cc. Since the diagnostic line
ratios we will use are almost entirely dependent on the abundances,
models adopting a different density will not change our results.  

As the metallicity of a photoionized nebula increases, the temperature
of the nebula decreases (as a result of the increased metal cooling),
and the relative strengths between hydrogen lines and lines from other
heavier elements also decreases (due to the increased abundance of the
heavy elements relative to hydrogen). The latter effect is especially
strong for nitrogen lines since the element is dominated by ``secondary"
production\footnote{where nitrogen is synthesized from existing carbon
and oxygen \citep{Pag81}.} at $Z \gtrsim$ 0.2 \zsun. Hence, N abundance
increases at a much faster rate with metallicity, \ie\ N/H $\propto
Z^2$. In Figure~\ref{fig:z}{\it a} we use the temperature sensitive
\othree\,$\lambda4363/\lambda5007$ ratio to demonstrate the dependance
of nebula temperature on metallicity. The stratification among the grids
of different metallicities is further improved by the fact that helium
increases linearly with metallicity relative to hydrogen \citep{Pag92}.
In Fig.~\ref{fig:z}{\it b} and Fig.~\ref{fig:n2}, we show the strong dependance of
\none\,$\lambda5199$/H$\beta$ and \ntwo\,$\lambda6584$/H$\alpha$ on metallicity. 
At each metallicity, we
show models spanning a range of ionization parameters ($-2.3 \leq$
log($U$) $\leq$ 0) and four power-law indices from $\alpha = -1.2$ to
$-2.0$, representing the quasar ionizing continuum ($F_{\nu} \propto
\nu^{\alpha}$). 
We have converted the modeled total metallicities to gas-phase
metallicities using the solar abundances defined by \citet{And89}
(12+log(O/H)$_{\odot}$ = 8.93 and \zsun\ = 0.02), in order to be
consistent with both previous studies on the metallicity of quasar BLRs
and metallicity results from stellar populations (\S~\ref{sec:spop}).
One solar metallicity in \citet{Gro04a} corresponds to 1/3 \zsun\ in our
figures, since approximately half of the metals are assumed to be
depleted onto dust in the models, and their assumed solar metallicity is
about 0.2 dex\footnote{The C, N, O, and Fe abundances in the solar
abundance set assumed by \citet{Gro04a} are all $\sim$0.2 dex lower than
those in \citet{And89}. Hence, the mass fraction of metals (i.e. $Z$) is
about 1.5 times lower --- \zsun\ = 0.013 --- than that of
\citet{And89}.} lower than that of \citet{And89}. 

As can be seen in Figs.~\ref{fig:z} \& \ref{fig:n2}, the line-ratios of the EELR as well
as the NLR of 3C\,79, like those quasar EELRs and their BLRs, are most
consistent with a low-metallicity at about 1/3 to 2/3 solar. 
Fig.~\ref{fig:n2} illustrates the peculiarity of 3C\,79 more clearly by comparing it
directly with SDSS AGN on the same line-ratio diagnostic. The nuclear NLR along with the two other 
regions where we have \ntwo\ measurements all lie on the lower metallicity side of the main Seyfert
branch, overlapping with quasar EELRs.
This result supports the orientation-based unification schemes of FR\,II radio galaxies and radio-loud quasars
\citep{Bar89}. On the other hand, 3C\,79 offers yet another example
strengthening (1) the correlation between the presence of EELRs and the
metallicity of the nuclear gas (BLRs and/or NLRs), and (2) the similar
metallicity in EELRs and the gas in the nuclear region. Both correlations 
were initially discovered only among steep-spectrum radio-loud quasars \citep{Fu07a}.  

\section{Discussion} \label{sec:dis}

\subsection{The Origin of the Low-Metallicity Gas}

The great majority of quasars show super-solar metallicities in their
nuclear regions \citep[see][for a recent review]{Ham07}; yet, as we
have previously shown \citep{Fu07a}, quasars with luminous EELRs are
drawn exclusively from the subset of steep-radio-spectrum quasars that
have BLRs with sub-solar metallicities.  The fact that the
quasar host galaxies are almost always very massive and thus expected to
have high metallicities suggests an external source for the
low-metallicity gas.  Furthermore, the apparent link between the
metallicity of the gas in the BLRs and the metallicity of
the gas in the EELRs, which have minimum masses of $10^{\rm9-10}$ \msun,
means that the external source of the gas must itself have been
moderately massive.

In 3C\,79, we have an FR\,II radio galaxy that shows the same pattern.
Although we of course cannot obtain a broad-line-region metallicity in
this case, we have used the unresolved nuclear narrow-line region
metallicity as a surrogate.  Aside from providing a consistency check on
unified models for FR\,II radio galaxies and quasars, the main question
of interest is whether the clearer view of the inner region of the host
galaxy gives us any insight into the origin of the low-metallicity gas
and the mechanism that produced the EELR.

The host galaxy of 3C\,79 appears to be a fairly normal elliptical
morphologically. The most intriguing feature in the inner region is the
extremely compact galaxy 0\farcs8 from the center of the host galaxy.
The nearly exponential profile of this galaxy suggests that it could be
a tidally stripped ``pseudobulge'' of a late-type galaxy
\citep[e.g.,][]{Kor04}; the interstellar medium of such a galaxy would
be quite plausible as a source for the low-metallicity gas in 3C\,79.
We have suggested elsewhere \citep{Sto07,Fu07b} that luminous EELRs may
result from nearly spherical blast waves connected with the initiation
of FR\,II radio jets.  The connection with low-metallicity gas is less
clear, although we can speculate that it may have something to do with
the lower radiative coupling such gas would have to the quasar radiation
field, allowing more efficient accretion.  It is interesting to note
that related considerations have been invoked recently to explain the link
between low metallicity and long-duration gamma-ray bursts \citep{Fru06}
on the scale of massive stars, although the physical mechanisms involved
are certainly different.

\subsection{Low-Metallicity Radio Galaxies in SDSS}

Armed with the latest photoionization models, \citet{Gro06} identified
$\sim$40 (out of $\sim23000$) candidates of low-metallicity Seyfert2s in
SDSS. A caveat to note is that the authors restricted the sample to
galaxies with stellar masses lower than $10^{10}$ \msun\  while
selecting their candidates, as guided by the mass-metallicity
correlation \citep[e.g.][]{Tre04}.  Although the EELR quasars are
excluded from the SDSS emission-line galaxy sample, 3C\,79 represents a
population of low-metallicity AGN that were missed by \citeauthor{Gro06}
These AGN are hosted by massive evolved galaxies with $\sim10^{12}$
\msun\ of stars and harbor $\sim10^9$ \msun\ black holes at their
hearts, and they are likely to be more powerful than the Seyfert2s.

By cross-correlating the TEXAS radio galaxies \citep{Dou96} with SDSS
DR4 AGN, we have identified a sample of low-metallicity radio galaxies.
Like that of 3C\,79, the spectra of their host galaxies show red slopes
and absorption features that are indicative of an old stellar
population. Follow-up high-resolution imaging and spatially resolved
spectroscopy of these galaxies and a comparable control sample are
needed to finally nail down the source of the low-metallicity gas and
the triggering mechanism for quasar superwinds. In addition, detecting
EELRs around these objects would provide a new test for the unification
schemes.

\section{Summary} \label{sec:sum}

Based on extensive ground-based spectroscopy and archival \hst\ WFPC2 images of
the radio galaxy 3C\,79, we have conducted a detailed analysis of its
host galaxy and the EELR. The host galaxy of 3C\,79 is a massive
elliptical with $M_R = -23.2$. The UV/optical spectral energy distribution of the host galaxy 
is best described by a combination of an intermediate-age stellar population (1.3 Gyr), 
an old stellar population (10 Gyr), a power law, and a nebular thermal continuum. Both stellar populations are metal rich (2.5 \zsun). This best fit model indicates a total stellar mass within one effective radius of approximately $2.5\times10^{11}$ \msun, consistent with the virial mass derived from stellar kinematics ($\sim4\times10^{11}$ \msun). 

The EELR of 3C\,79 is remarkably similar to the most luminous quasar EELRs. The velocity
field, although only available in a small area, is locally ordered but
globally disordered. The EELR is almost certainly photoionized by the
hidden quasar, and it shows densities of $\sim100$ \cc\ in the \otwo\
emitting regions and temperatures around 13,000 K. Most interestingly,
the metallicity of the gas in both the EELR and the NLR is
about 1/3$-$2/3 solar, matching perfectly the metallicity in both the EELRs and 
the nuclear BLRs of the EELR quasars.  

There is a very compact close companion galaxy 3.3 kpc from and 4 magnitudes fainter than the host galaxy. This companion galaxy shows an exponential profile, and 
is presumably a tidally stripped ``pseudobulge" of a late-type galaxy that might be the source for the low-metallicity gas as well as the starforming regions in 3C\,79. 

The exact correspondence between this EELR and the EELRs around
quasars joins the already overwhelming evidence in support of the
unification schemes for FR\,II radio galaxies and radio-loud quasars.
The definitive trait of sub-solar metallicity in the NLR of 3C\,79 also
provides an efficient tool for selecting FR\,II radio galaxies that
likely host luminous EELRs.

\acknowledgments 
We thank the Gemini staff for carrying out the GMOS observations and Nicola Bennert for helping us obtain the LRIS spectra. We also thank the anonymous referee for a careful reading of the manuscript and for cogent comments to help us clarify and improve the presentation. This research has been partially supported by NSF grant AST03-07335. 


\clearpage
\begin{landscape}

\begin{deluxetable}{lccccccc}
\tablewidth{0pt}
\tabletypesize{\scriptsize}
\tablecaption{Morphology of 3C\,79 Host Galaxy and Its Close Companion \label{tab:morph}}
\tablehead{
\colhead{Galaxy} & \colhead{Model} & \colhead{F675W$^a$} & \colhead{F702W$^a$} & \colhead{$r_{1/2}$/kpc}& \colhead{$n^b$} & \colhead{b/a} & \colhead{P.A.}
}
\startdata
3C79A   & \sersic\ & \nodata & 22.19   & 0.22   &  1.23  &  0.49 &  18.2 \\
\nodata  & Exp & 22.24 & 22.23  & 0.22  & (1) &  0.48 &  18.0 \\
3C79     & \sersic\ & 18.06  & \nodata & 10.1  &  5.21 &  0.70 &  6.2 \\
\nodata  & de Vauc & 18.24 & \nodata & 7.2 & (4) &  0.71 &   5.7 \\
\hline
Dunlop$^c$ & de Vauc & 17.46 & \nodata & 7.5 & (4) & 1.0 & 13 
\enddata
\tablenotetext{a}{ST magnitudes corrected for Galactic extinction.}
\tablenotetext{b}{\sersic\ indexes.}
\tablenotetext{c}{Converted from \citet{Dun03} Table 3 to our cosmology.}
\end{deluxetable}

\begin{deluxetable}{lcccccccccc}
\rotate
\tablewidth{0pt}
\tabletypesize{\scriptsize}
\tablecaption{Line Ratios of 3C\,79 Line-Emitting Clouds Relative to H$\beta$
\label{tab:flux}}
\tablehead{
\colhead{Region}
& \colhead{[Ne\,{\sc V}]\,$\lambda3426$} & \colhead{[O\,{\sc II}]\,$\lambda3727$} & \colhead{[Ne\,{\sc III}]\,$\lambda3869$}
& \colhead{[O\,{\sc III}]\,$\lambda4363$} & \colhead{He\,{\sc II}\,$\lambda4686$} & \colhead{H$\beta$}
& \colhead{[O\,{\sc III}]\,$\lambda5007$} & \colhead{[N\,{\sc I}]\,$\lambda5199$}
& \colhead{H$\alpha$} & \colhead{[N\,{\sc II}]\,$\lambda6584$}
}
\startdata
\multicolumn{11}{c}{Gemini GMOS/IFU} \\
\hline
  $a$ & $0.83\pm0.19$ & $2.45\pm0.16$ & $0.65\pm0.07$ & $0.15\pm0.05$ & $0.33\pm0.04$ & $1.00\pm0.05$ & $9.32\pm0.03$ & $<0.18$ & \nodata & \nodata \\
  $b$ & $1.54\pm0.43$ & $3.25\pm0.35$ & $1.18\pm0.19$ & $<0.49$ & $<0.33$ & $1.00\pm0.11$ & $8.91\pm0.08$ & $<0.47$ & \nodata & \nodata \\
  $c$ & $0.53\pm0.10$ & $2.50\pm0.16$ & $1.01\pm0.05$ & $0.13\pm0.03$ & $0.26\pm0.02$ & $1.00\pm0.02$ & $10.39\pm0.02$ & $0.06\pm0.02$ & \nodata & \nodata \\
  $d$ & $1.38\pm0.15$ & $2.62\pm0.09$ & $1.19\pm0.05$ & $0.19\pm0.02$ & $0.29\pm0.02$ & $1.00\pm0.02$ & $11.52\pm0.01$ & $0.05\pm0.02$ & \nodata & \nodata \\
  $NLR$ & $1.057\pm0.016$ & $1.590\pm0.015$ & $0.944\pm0.004$ & $0.176\pm0.003$ & $0.302\pm0.003$ & $1.000\pm0.004$ & $10.923\pm0.002$ & $0.041\pm0.004$ & \nodata & \nodata \\
\hline
\multicolumn{11}{c}{Keck LRIS/Long-Slit} \\
\hline
$N1$ & $0.60\pm0.02$ & $2.73\pm0.02$ & $1.09\pm0.02$ & $0.15\pm0.03$ & $0.28\pm0.02$ & $1.00\pm0.02$ & $10.63\pm0.02$ & $<0.09$ & $3.09\pm0.03$ & $1.24\pm0.03$ \\
$S1$ & $0.58\pm0.04$ & $2.92\pm0.05$ & $1.31\pm0.04$ & $0.27\pm0.07$ & $<0.15$ & $1.00\pm0.04$ & $11.83\pm0.03$ & $<0.21$ & $2.93\pm0.07$ & $0.91\pm0.06$ \\  $NLR$ & $1.093\pm0.003$ & $1.591\pm0.015$ & $0.921\pm0.009$ & $0.185\pm0.008$ & $0.297\pm0.007$ & $1.000\pm0.008$ & $10.602\pm0.008$ & $0.038\pm0.017$ & $3.100\pm0.008$ & $1.040\pm0.007$ 
\enddata
\end{deluxetable}

\begin{deluxetable}{lcccccccc}
\rotate
\tablewidth{0pt}
\tabletypesize{\scriptsize}
\tablecaption{Properties of 3C\,79 Line-Emitting Clouds
\label{tab:prop}}
\tablehead{
\colhead{Region}
& \colhead{$V$}
& \colhead{$\sigma$} 
& \colhead{$A_V$\tablenotemark{a}}
& \colhead{H$\beta \times 10^{17}$}
& \colhead{[O\,{\sc ii}]}
& \colhead{[O\,{\sc iii}]}
& \colhead{$N_e$}
& \colhead{$T_e$}
\\
\colhead{}
& \colhead{(\kms)}
& \colhead{(\kms)}
& \colhead{(mag)}
& \colhead{(erg cm$^{-2}$ s$^{-1}$)}
& \colhead{3726/3729}
& \colhead{(4959+5007)/4363}
& \colhead{(\cc)}
& \colhead{(K)}
}
\startdata
         $a$  &   $-145$  &    102  &   (0.000)  &     $4.7\pm0.2$  &    \nodata  &   $85\pm29$  & \nodata &   \nodata      \\
         $b$  &   $-138$  &     91  &   (0.000)  &     $1.4\pm0.2$  &    \nodata  &   $>23$  & \nodata &   \nodata      \\
         $c$  &    $-74$  &    109  &   0.016  &     $21.0\pm0.4$  &    $0.76\pm0.08$  &  $108\pm22$  &     $100\pm100$ &  $12500^{+1200}_{-800}$ \\
         $d$  &   $+131$  &    132  &  0.392  &    $23.5\pm0.4$  &    $0.80\pm0.08$  &  $82\pm10$  &     $150\pm100$ &   $13900^{+800}_{-700}$ \\
         $NLR$  &     $0$  &    136  &   0.284  &   $99.6\pm0.4$  &    $0.70\pm0.07$  &   $83.0\pm1.4$  &      $30\pm80$ &   $13880\pm100$ 
\enddata
\tablenotetext{a}{Intrinsic reddening. The values in parentheses are not directly measured from the H$\gamma$/H$\beta$ ratio, since the H$\gamma$ lines in these
spectra are not well-detected; and we assumed zero reddening for these clouds. 
See \S~\ref{sec:ne_t} for details.}
\end{deluxetable}
\clearpage
\end{landscape}

\clearpage

\end{document}